\documentclass[letter]{aa}  

\usepackage{graphicx}
\usepackage{txfonts}
\usepackage{amssymb}
\usepackage{multirow}
\usepackage{float}
\usepackage{gensymb}
\usepackage{xcolor}
\usepackage{verbatim} 
\usepackage{rotating}
\usepackage{sidecap}
\usepackage[]{hyperref}
\hypersetup{backref=true, pagebackref=true, hyperindex=true, breaklinks=true,colorlinks=true,urlcolor=blue, linkcolor=blue, citecolor=blue,pagecolor=red, bookmarks=true, bookmarksopen=true}

\begin{document}

    \title{ALMA chemical survey of disk-outflow sources in Taurus (ALMA-DOT)}

   \subtitle{II: Vertical stratification of CO, CS, CN, H$_2$CO, and CH$_3$OH in a Class I disk} 
   
%   III: The vertical disk chemical structure of the butterfly star IRAS 04302+2247}

%   \subtitle{III: Vertical distribution of molecular emission around the butterfly star}

   \author{L. Podio \inst{1}
   \and
          A. Garufi \inst{1}
          \and
          C. Codella \inst{1, 2}
          \and
          D. Fedele \inst{1}
          \and
          E. Bianchi
          \inst{2}
          \and
          F. Bacciotti           
          \inst{1}
          \and
          C. Ceccarelli
          \inst{2}
          \and
          C. Favre
          \inst{2}
          \and
          S. Mercimek
          \inst{1, 4}
          \and
          K. Rygl
          \inst{3}
          \and
          L. Testi          
          \inst{5, 6, 1}
          }

   \institute{INAF - Osservatorio Astrofisico di Arcetri, Largo E. Fermi 5, 50125 Firenze, Italy\\
              \email{lpodio@arcetri.astro.it}
         \and Univ. Grenoble Alpes, CNRS, IPAG, 38000 Grenoble, France
         \and INAF - Istituto di Radioastronomia \& Italian ALMA Regional Centre, via P. Gobetti 101, 40129 Bologna, Italy
         \and Universit{\`a} degli Studi di Firenze, Dipartimento di Fisica e Astronomia, Via G. Sansone 1, 50019 Sesto Fiorentino, Italy
         \and European Southern Observatory, Karl-Schwarzschild-Strasse 2, 85748 Garching bei M{\"u}nchen, Germany
         \and Excellence Cluster Origins, Boltzmannstrasse 2, 85748 Garching bei M{\"u}nchen, Germany
             }

   \date{Received -; accepted -}

% \abstract{}{}{}{}{} 
% 5 {} token are mandatory
 
  \abstract
   {The chemical composition of planets is inherited from that of the natal protoplanetary disk at the time of planet formation. Increasing observational evidence suggests that planet formation occurs in less than 1$-2$ Myr. This motivates the need for spatially resolved spectral observations of young Class I disks, as carried out by the ALMA chemical survey of Disk-Outflow sources in Taurus (ALMA-DOT). In the context of ALMA-DOT, we observe the edge-on  disk around the Class I source IRAS 04302+2247 (the butterfly star) in the 1.3~mm continuum and five molecular lines. We report the first tentative detection of methanol (CH$_3$OH) in a Class I disk and resolve, for the first time, the vertical structure of a disk with multiple molecular tracers. The bulk of the emission in the CO $2-1$, CS $5-4$, and o-H$_2$CO $3_{1,2}-2_{1,1}$ lines originates from the warm molecular layer, with  the line intensity peaking at increasing disk heights, $z$, for increasing radial distances, $r$. Molecular emission is vertically stratified, with CO observed at larger disk heights (aperture $z/r \sim 0.41-0.45$) compared to both CS and H$_2$CO, which are nearly cospatial ($z/r \sim 0.21-0.28$). In the outer midplane, the line emission decreases due to molecular freeze-out onto dust grains (freeze-out layer) by a factor of $>100$ (CO) and $15$ (CS). The H$_2$CO emission decreases by a factor of only about $2$, which is possibly due to H$_2$CO formation on icy grains, followed by a nonthermal release into the gas phase.
%   We also constrain 
    The inferred [CH$_3$OH]/[H$_2$CO] abundance ratio is $0.5-0.6$, which is $1-2$ orders of magnitude lower than for Class 0 hot corinos, and a factor  $\sim2.5$ lower than the only other value inferred for a protoplanetary disk (in TW Hya, $1.3-1.7$). Additionally, it is at the lower edge but still consistent with the values in comets. This may indicate that some chemical reprocessing occurs in disks before the formation of planets and comets.  %Resolved observations of molecular emission in disks from the Class 0 to the Class II stages are needed to infer reliable abundance ratios of key organics and establish the chemical evolution from the protostellar stage to planet-forming disks.
   }

   \keywords{Protoplanetary disks -- Astrochemistry -- ISM: molecules -- Stars: individual: IRAS 04302+2247}

\authorrunning{Podio et al.}

\titlerunning{Chemical structure of the disk of IRAS 04302+2247}

   \maketitle
%
%-------------------------------------------------------------------

\section{Introduction}

The chemical composition of exoplanets is determined by that of their natal environment, the protoplanetary disk. Recent studies suggest that planet formation occurs earlier than previously thought, that is, in disks of less than 1 Myr, which are massive enough to form planets with the observed exoplanetary masses \citep{tychoniec20}. Furthermore, these young disks show gaps in their dust grains distribution, which is a possible signature of forming planets \citep[e.g.,][]{fedele18,sheehan17,sheehan18}. 
It is therefore crucial to probe the chemical composition of young disks around Class I sources ($< 10^6$ years) in order to determine the following: what molecules are present in young disks and how they are distributed; whether disks inherit the chemical complexity observed at the protostellar stage \citep[e.g., ][]{lee19a,jorgensen16} in terms of complex organic (COMs) and prebiotic molecules, as suggested by the recent work of \citet{bianchi19a} and \citet{drozdovskaya19}; and whether COMs are formed in the disk due to efficient ice chemistry in the cold midplane \citep[e.g., ][]{walsh14}.

According to thermo-chemical models, protoplanetary disks consist of three chemical layers \citep[e.g., ][and references therein]{aikawa02,dullemond07,dutrey14}:
(i) the hot surface layer, or disk atmosphere, where molecules are photodissociated; 
%and only atoms, ions, radicals, and PAHs are able to survive;
(ii) the warm molecular layer, where molecules are in the gas phase and gas-phase chemistry is at play;
%efficiently desorbed from grains due to non-thermal processes, and (re-)formed through gas-phase reactions;
and (iii) the freeze-out layer, that is, the cold outer disk midplane where molecules freeze out onto dust grains. 
For each molecule, the freeze-out occurs at the disk radius and height where the dust temperature falls below the freeze-out temperature, which depends on the molecular binding energy.
 %when the dust temperature falls below their binding energy.  %where those molecules with binding energy higher than the dust temperature freeze out on dust grains.
% These three layers are stratified both vertically and radially.%
 %The freeze-out of molecules depends on their binding energy. 
% The dust temperature decreases both at larger distance from the star and deeper height in the disk. Thus, the snowlines of molecules with decreasing binding energies are displaced radially at increasing radii and vertically at increasing disk heights.
%The position of the snowlines is determined by the molecules binding energy. 
However, frozen-out molecules outside their snowline can be released into the gas via nonthermal desorption processes \citep[e.g.,][and references therein]{willacy09,walsh14,loomis15}.
 These processes are efficient in the molecular layer, that is, in the inner disk region and at increasing disk heights in the outer disk, due to (inter-)stellar UV and X-rays penetration. In contrast, molecules are only partially released into the gas phase in the outer midplane mostly due to reactive and cosmic-ray-induced desorption \citep[e.g.,][and references therein]{walsh14}. 
 %The molecular layer comprehends theinner  disk  region  and  then its height above the disk midplane increases radially. 
% Thus the bulk of molecular emission is expected to originate from this layer. 
 As a result, the chemical composition of the ices in the disk midplane remains largely hidden to observations. However, recent observations and modeling suggest that the relatively high gas-phase abundance and intensity of formaldehyde (H$_2$CO) in the outer disk region can be explained by formation on, and a release from, the icy grains in the disk midplane \citep[e.g.,][and references therein]{carney17,kastner18,loomis15,oberg17,podio19,pegues20}.
% due to CO hydrogenation outside the CO snowline
  A key molecule to test the chemical composition of ices in the midplane, where planets form, is methanol (CH$_3$OH). This molecule is one of the building blocks for the formation of more complex organic molecules and, as opposed to formaldehyde, CH$_3$OH only forms on the icy mantles of dust grains \citep{watanabe02}. 
  %Methanol can be either formed in the prestellar and protostellar phase, then inherited by the disk at the time of its formation (inheritance scenario); and/or it can form in the disk midplane (disk reset). In both cases,
  According to disk models, CH$_3$OH 
  %stored on the ices in the midplane and released via non-thermal processes 
  can reach gas-phase abundances of up to $10^{-8}$  \citep[e.g., ][]{walsh14}. However, due to its large partition function, to date, methanol has only been detected in the disks of TW Hya \citep{walsh16} and of the young outbursting star V883 Ori \citep{vanthoff18,lee-je19}.

\begin{figure*}
  \centering
 \includegraphics[width=18cm]{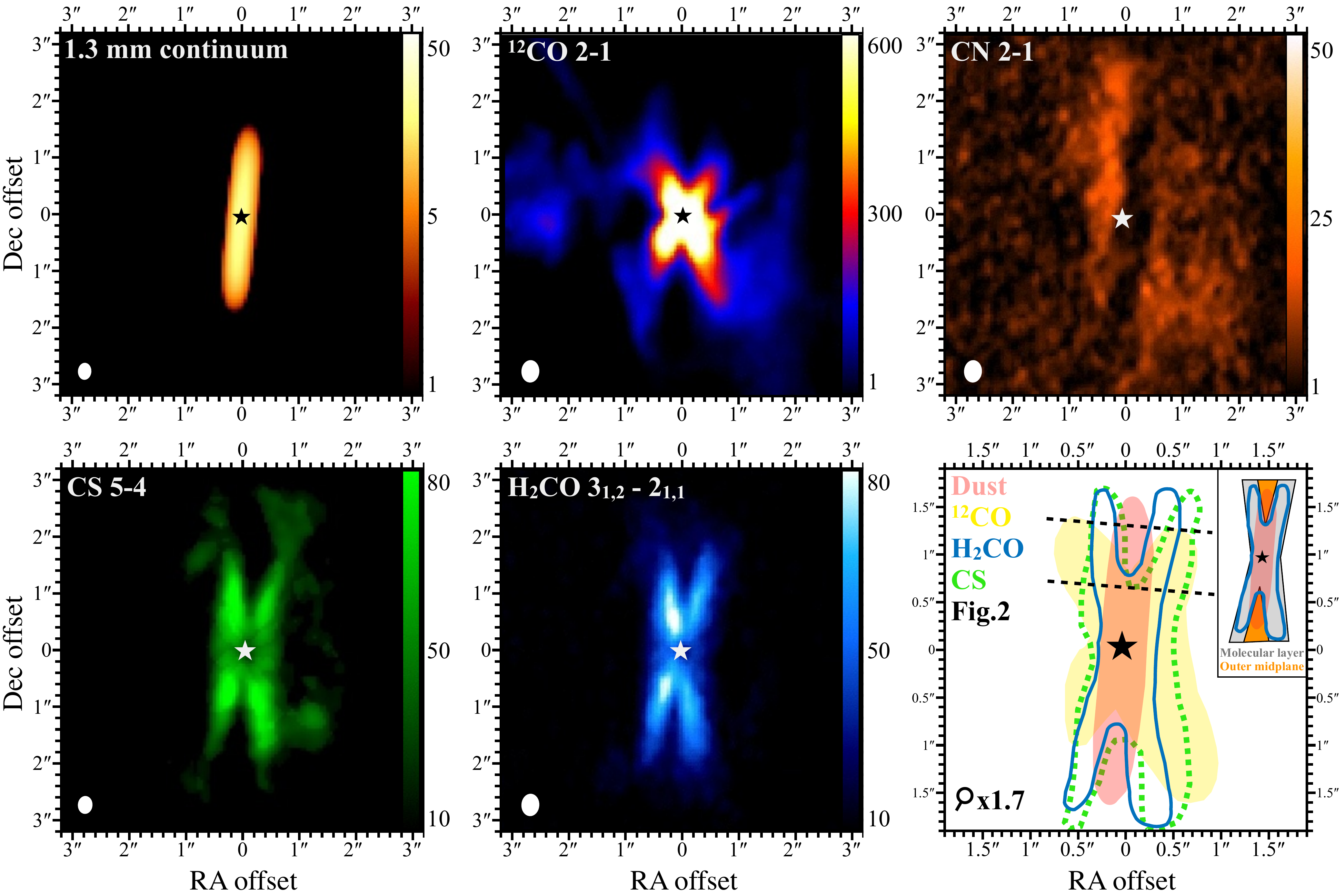}
     \caption{Moment 0 maps of continuum at 1.3~mm and molecular emission toward IRAS 04302+2247. Molecular lines are the CO $2-1$, CN $2-1$, CS $5-4$, and H$_2$CO $3_{1,2}-2_{1,1}$. In each panel: The color wedge on the right shows the intensity in units of mJy beam$^{-1}$ and in logarithmic scale for the continuum, and in units of mJy beam$^{-1}$ km s$^{-1}$ and linear scale for the lines. The star in the center indicates the geometrical center of the continuum emission. The beam size is shown at the bottom left. The sketch in the bottom-right panel shows the comparison between the continuum (dust) and molecular emission, and it is zoomed in by a factor of 1.7. The black-dashed lines indicate the section over which the vertical profiles of Fig.\,\ref{fig:vertical-profiles} are obtained. The inset shows the regions named the molecular layer and outer midplane (see Sect.\,\ref{column_densities}). North is up, east is left.} 
 \label{fig:overview}
 \end{figure*}

%to constrain the thermal and chemical vertical structure of the disk to verify thermochemical models and constrain the gas phase abundances of molecules in young disks
  The most direct way to investigate the distribution and origin of molecules
  %(formation and release mechanisms)
  is to observe the vertical structure of disks so as to resolve the distinct chemical layers. The vast majority of the available resolved observations of molecular emission from protoplanetary disks, however, is relative to disks seen at low or medium inclinations ($i<75\degree$). These observations only directly probe the disk radial structure, but not the vertical one. 
  %However, resolving the disk vertical extent is crucial to understand the formation mechanism of molecules, either in gas-phase in the warm molecular layer and/or on the surface of grains in the cold disk-midplane, and if there is a strong contributions to the emission from the cold freeze-out layer. 
  To date, resolved observations of edge-on disks have only been obtained in CO and its isotoplogues \citep[e.g., HH30 and IRAS 18059-3211,][]{louvet18, teague20}.
  
  An ideal target to investigate the disk chemical structure is the edge-on disk
  ($i = 90\pm3\degree$\footnote{A a lower inclination, $i \sim 76\degree$ was estimated by \citet{sheehan17b}},  \citealt{wolf03})
  %($70\degree-90\degree$,  \citealt{sheehan17b}) 
  around the Class I source IRAS 04302+2247 (hereafter IRAS04302), also known as the butterfly star \citep{lucas97} and located in Taurus ($d=161\pm3$ pc, \citealt{galli19}). This source was observed 
  %in the context of the disk surveys performed
  with the IRAM-30m in CO, H$_2$CO, CS, SO, HCO$^+$, and HCN lines  \citep{guilloteau13,guilloteau16}, and with the PdBI in the continuum  \citep{guilloteau11}. Based on their double-peaked profile, H$_2$CO and SO emission is attributed to the disk, while scattered-light images reveal the prominent outflow cavity \citep{lucas97, Padgett1999, Eisner2005}. %\citet{Wolf2008} revealed a flux minimum in the center at 0.9 mm ascribed to an increased optical depth but this dip is not visible at 1.3 mm \citep{Grafe2013}.
%  In this work, we report on new images at $\sim 0\farcs3$ from the Atacama Large Millimeter Array (ALMA) of the well-known, edge-on disk around the Class I source IRAS 04302+2247 \citep[hereafter IRAS 04302 but also known as the butterfly star,][]{lucas97}. This source located in the Taurus star forming region ($d=140$ pc) was previously observed in the context of the disk surveys performed with the IRAM-30m in CO, H$_2$CO, CS, SO, HCO$^+$, and HCN lines  \citep{guilloteau13,guilloteau16}, and with the PdBI in the continuum  \citep{guilloteau11}. Based on their double-peaked profile H$_2$CO and SO emission is attributed to the disk, while scattered-light images reveal the prominent outflow cavity \citep{lucas97, Padgett1999, Eisner2005}. %\citet{Wolf2008} revealed a flux minimum in the center at 0.9 mm ascribed to an increased optical depth but this dip is not visible at 1.3 mm \citep{Grafe2013}.
In this work, we report on new images of IRAS 04302 at $\sim 0\farcs3$ ($\sim 48$ au) taken with the Atacama Large Millimeter/submillimeter Array (ALMA) in the context of the ALMA-DOT program (ALMA chemical survey of Disk-Outflow sources in Taurus, see \citealt{garufi20b}, and Garufi et al. in preparation). 
%The emission in several molecular tracers (CO, H$_2$CO, CS, CN, and CH$_3$OH) reveal the disk chemical structure and allows us to derive stringent constraints on the molecules abundance and distribution. 
%has the goal to spatially disentangle the disk, outflow, and medium emission in young, embedded sources as well as to broaden the chemical characterization of disks by imaging in parallel nineteen different lines of nine different species. In the context of ALMA-DOT, we observed IRAS 04302 with $\sim$0.3\arcsec resolution resolving the molecular emission of CO, H$_2$CO, CS, CN, and CH$_3$OH, thus as to obtain the chemically richest dataset ever obtained for an edge-on disk \textbf{(true?)}.

%The origin of the CO, CN, HCO$^{+}$, and HCN emission is more unclear due to possible contamination from the surrounding envelope and outflow.

%--------------------------------------------------------------------
\section{Observations and data reduction} \label{data_reduction}

Observations were taken with ALMA-Band 6 on October 28, 2018  with baselines ranging from 15~m to 1.4 km (project 2018.1.01037.S, PI: L. Podio). The integration time sums up to $\sim 113$ minutes. The bandpass and phase calibrators are J0423-0120 and J0510+1800, respectively. 
The correlator set-up consists of high-resolution (0.141 MHz) spectral windows (SPWs), covering CO 2$-1$,  CN $2-1$, o-H$_2$CO $3_{1,2}-2_{1,1}$, CS $5-4$,  and CH$_3$OH $5_{0,5}-4_{0,4}$ (A). The lines' properties are summarized in Table \ref{tab:lines}.
%\footnote{From the Cologne Database of Molecular Spectroscopy \citep[CDMS,][]{Mueller2005}}.
Data reduction was carried out using CASA 4.7.2.
Self-calibration was performed on the continuum emission and applied on the line-free continuum and continuum-subtracted line emission. The signal-to-noise ratio (S/N) of the continuum improved by a factor of 3.4 after the self-calibration. The final maps were produced with \textsc{tclean} by applying a manually selected mask on the visible signal. We used Briggs weighting with robust=0.0 for the bright CO, H$_2$CO, CS, and CN lines to obtain high angular resolution maps, while we set Briggs weighting with robust=2.0 for the faint CH$_3$OH line to maximize the S/N at the expense of angular resolution. The channel width is 0.2 km s$^{-1}$, except for the CS line (1.2 km s$^{-1}$), which is covered by the broad SPW for the continuum (1.129 MHz resolution). The clean beam of the self-calibrated maps ranges from 0.31\arcsec$\times$0.26\arcsec\ to 0.41\arcsec$\times$0.32\arcsec\ and the root mean square (r.m.s.) noise per channel is $\simeq 0.8-2.3$ mJy beam$^{-1}$. Moment 0 and 1 maps were produced over the velocity range [$+1.2, +10$] km s$^{-1}$, except for CO ([$-3$, $+14.2$] km s$^{-1}$). 
%Only fluxes above $3\sigma$ are included in the calculation of the Moment 1 maps.   

\section{Results}

Figure \ref{fig:overview} shows the moment 0 maps of the continnum at 1.3~mm and molecular emission toward IRAS 04302, while moment 1 maps are shown in Fig.~\ref{fig:mom1}. The continuum emission reveals the silhouette from the edge-on disk, which is roughly oriented along the north-south direction (P.A.=175$\degree$). The half-width of the emission along the disk's minor axis is $\sim$0.5\arcsec, which is more than twice the beam size along this direction (0.22\arcsec). Thus, the disk's vertical extent is resolved. The distribution of the line emission is different across the various molecules both radially (i.e., across the disk radial extent) and vertically (i.e., across the disk height), as is discussed in the following sections.

\subsection{Vertical distribution of molecular emission}

The emission from CO $2-1$, CS $5-4$, and o-H$_2$CO $3_{1,2}-2_{1,1}$ shows an X-shaped structure, that is, the line intensity peaks at increasing disk heights for increasing radial distances out to $\sim3$\arcsec\ ($\sim 480$ au). The CO emission is centrally-peaked, while the CS and H$_2$CO emission are dimmed in the inner 0.3\arcsec\ ($\sim 48$ au). As opposed to the other molecular tracers, the CN $2-1$ emission does not show an X-shape and is instead detected at a constant height above the midplane across the disk's entire radial extent. Negative values are seen in the region corresponding to the 1.3~mm continuum emission, that is, the dusty disk is seen in silhouette. This is caused by continuum over-subtraction, which is likely due to the absorption of the continuum by interstellar and circumstellar CN molecules along the line of sight. % at the CN $2-1$ line rest frequency.

\begin{figure}
  \centering
 \includegraphics[width=9cm]{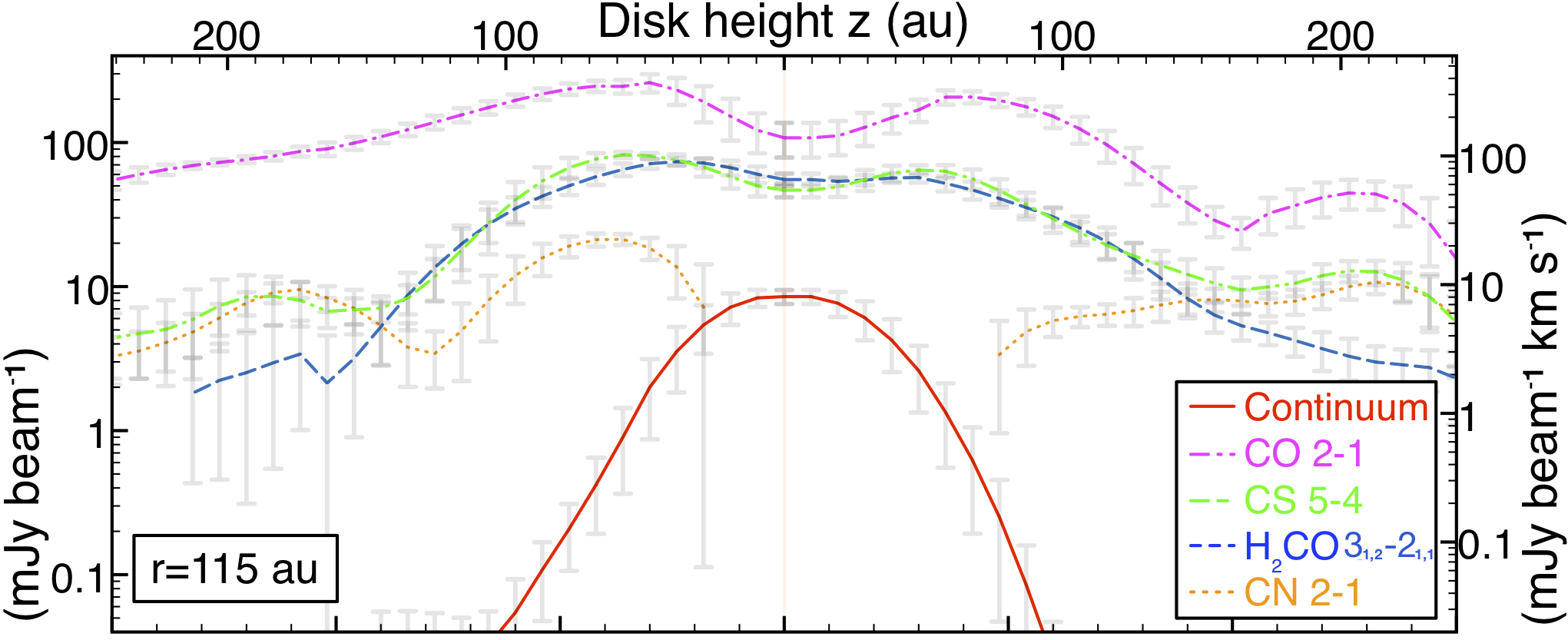}
  \includegraphics[width=9cm]{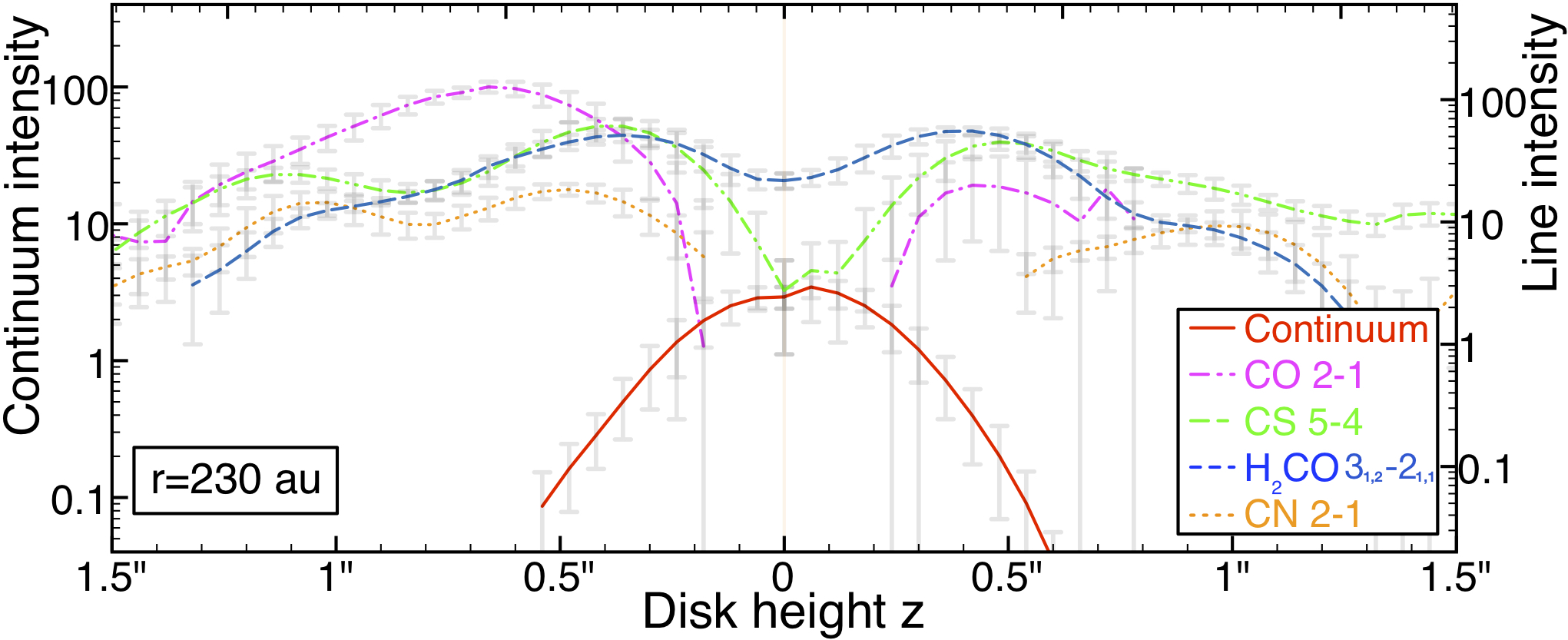}
 \caption{Vertical intensity profiles of CO $2-1$ (magenta), CS $5-4$ (green), H$_2$CO $3_{1,2}-2_{1,1}$ (blue), CN $2-1$ (brown) (in mJy beam$^{-1}$ km s$^{-1}$), and of the 1.3~mm continuum (in mJy beam$^{-1}$, red line). The vertical profiles were extracted at radial distances of 115~au (top) and 230~au (bottom).} 
 \label{fig:vertical-profiles}
 \end{figure}

A vertical stratification of the CO, CS, and H$_2$CO emission across the disk height is observed, as summarized in the sketch shown in the bottom-right panel of Fig.\,\ref{fig:overview}. The CO emission extends up to larger disk heights than the other molecules. The H$_2$CO and CS emission is cospatial on the east side, while the CS emission is slightly higher on the west side. To quantify the disk height $z$ from where the bulk of the emission in each molecule originates, we extracted the vertical profiles of the line intensity as a function of the radial distance. The profiles were extracted from the line moment 0 maps across the disk height, that is,  perpendicularly to the disk's major axis by averaging the emission radially over three pixels (0\farcs18, corresponding to $\sim 29$ au).  
The vertical profiles extracted to the north at radial distances of $115$ and $230$ au  are shown in Fig.\,\ref{fig:vertical-profiles}. The errors in the figure correspond to the standard deviation obtained over a nine-pixel box.
%From the vertical intensity profiles the disk height $z$ where the emission peak at each radii is derived. 
We find that the disk height $z$ where the emission peaks linearly increases with the radial distance $r$, resulting in a constant disk aperture $z/r$ for each molecule. The emission in the different tracers is stratified, as $z/r$ spans from 0.41$-$0.45 for CO to 0.22$-$0.28, and 0.21$-$0.25 for CS, and H$_2$CO, respectively. The given $z/r$ ranges include the scattering encountered along the four different disk surfaces, that is to say northeast, southeast, northwest, and southwest.

The molecular emission decreases in the outer disk midplane for $r >150$ au. %The emission from these three molecules also differ for the amount of flux detected at the midplane. 
%This can be appreciated from the vertical profiles, i.e.\ across the disk height at different radial distances (r=100 au and r=200 au, see Fig.\,\ref{fig:vertical-profiles}). These profiles were extracted from the moment 0 maps perpendicularly to the disk by averaging the emission over 3 pixels (0.18\arcsec). The errors reported in Fig.\,\ref{fig:vertical-profiles} correspond to the standard deviation obtained over a 9-pixel box.
The vertical profiles show that at a disk radius of $115$ au, the intensity of all molecules is only slightly lower at the disk midplane with respect to their peak intensity (by a factor of $3$ for CO, $2$ for CS, and less than 2 for H$_2$CO). At $r=230$ au instead, the emission in the midplane is lower with respect to the peak intensity by a factor of 2 for H$_2$CO, and by a factor of 15 for CS. No CO emission is detected at the midplane, which means that it is lower than at the intensity peak by a factor of more than 100.

\subsection{Tentative detection of methanol}
\label{sect:methanol}

\begin{figure*}
  \centering
 \includegraphics[width=18.5cm]{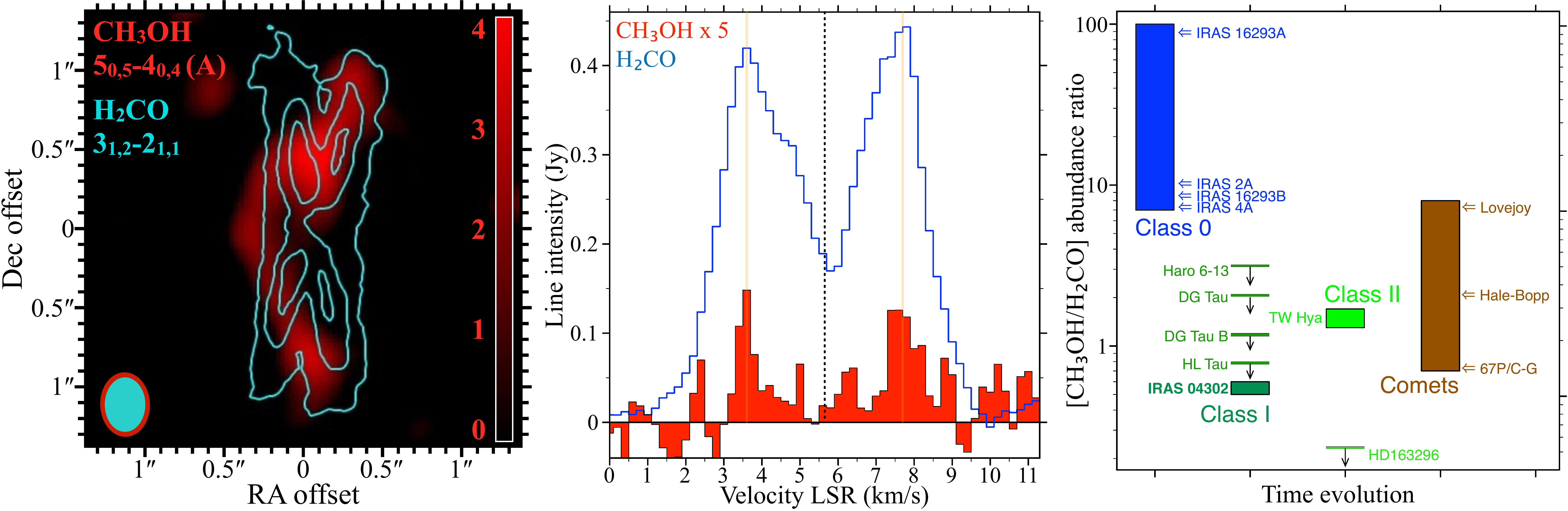} 
     \caption{{\it Left:} CH$_3$OH $5_{0,5}-4_{0,4}$ (A) moment 0 map. The color wedge shows the intensity in units of mJy beam$^{-1}$ km s$^{-1}$. The cyan contours indicate the H$_2$CO $3_{1,2}-2_{1,1}$ emission of Fig.\,\ref{fig:overview} at 20, 40, and 60 mJy beam $^{-1}$. The beam sizes are shown at the bottom left with the relative colors.  {\it Middle:} Spectral profiles of the H$_2$CO and CH$_3$OH lines integrated over the H$_2$CO emitting area. The CH$_3$OH profile is multiplied by 5. The solid-vertical lines indicate the peaks of the H$_2$CO profile and correspond to the peaks of the CH$_3$OH line. The dashed line indicates the source systemic velocity ($V_{\rm sys} = +5.6$ km s$^{-1}$). {\it Right:}  [CH$_3$OH]/[H$_2$CO] abundance ratio inferred for IRAS 04302 is compared with literature values inferred for Class 0 hot corinos, other Class I or early Class II disks observed by ALMA-DOT, Class II disks, and comets (references in the text).} 
 \label{fig:ch3oh}
 \end{figure*}

The moment 0 map of the CH$_3$OH $5_{0,5}-4_{0,4}$ (A) line shows emission up to 8 mJy beam$^{-1}$ km s$^{-1}$. The emission does not show the same spatial distribution as any of the other tracers shown in Fig. \ref{fig:overview}, but it is confined within the H$_2$CO emitting region, suggesting that it originates from the disk (see left panel of Fig.~\ref{fig:ch3oh}). 
%The CH$_3$OH kinematics is in agreement with that of the other molecular tracers. More specifically, 
This is further supported by the CH$_3$OH spectral profile obtained by integrating over the H$_2$CO emitting region (middle panel of Fig. \ref{fig:ch3oh}). This reveals two peaks that are symmetrically displaced at $\pm 2$ km s$^{-1}$ with respect to the systemic velocity (V$_{\rm sys} = +5.6$ km s$^{-1}$),
 which is %at $+3.6$ and $+7.6$ km s$^{-1}$,}
in perfect agreement with the peaks of H$_2$CO (middle panel of Fig.~\ref{fig:ch3oh}). The CH$_3$OH intensity obtained from the moment 0 map by integrating over the molecular layer, which is defined as the X-shaped region where CS and H$_2$CO emission is brighter (area $A=4.3$ arcsec$^2$, see sketch in Fig.\,\ref{fig:overview}), amounts to 42 mJy km s$^{-1}$. The noise on the integrated emission\footnote{\label{note1}The noise on the line integrated emission is $\sigma_0 \times \sqrt{A/\theta_{\rm beam}}$, where $\sigma_0$ is the r.m.s of the moment 0 map, $A$ is the area of integration, and $\theta_{\rm beam}$ is the beam size. The values of $\sigma_0$ and $\theta_{\rm beam}$ are listed in Table \ref{tab:lines}.} is 13 mJy km s$^{-1}$; therefore, this is a 3$\sigma$ detection. As an  a posteriori test, the integrated line intensity is consistent with the expectations for the [CH$_3$OH]/[H$_2$CO] abundance ratio (see Sect. 3.3).

%Weak CH$_3$OH flux is detected from the moment 0 map in correspondence of the gaseous disk (see left panel of Fig.\,\ref{fig:ch3oh}). Several evidence contribute to lend credence to such a detection. First of all, the kinematics of the flux is consistent with the disk rotation, being the flux to North blue and to South red (see Moment 1 map of H$_2$CO in the Appendix). Secondly, the CH$_3$OH spectral profile integrated over the H$_2$CO emitting region reveals two faint peaks that are perfectly coincident in velocity with the H$_2$CO peaks (see right panel of Fig.\,\ref{fig:ch3oh}). Finally, the CH$_3$OH flux integrated over this region amounts to 68 mJy km s$^{-1}$ and, since the noise expected over this area is 22 mJy km s$^{-1}$, we can firmly claim a 3$\sigma$ detection of methanol. As a posteriori test, this amount of flux is consistent with the expectations for the CH$_3$OH/H$_2$CO column density ratio (see Sect.\,\ref{column_densities}).  
%The moderate angular resolution of the available moment 0 map ($\sim$0.37\arcsec) and the faint nature of the line do not allow the spatial characterization of this emission. On the disk North side, where the line is slightly brighter, the CH$_3$OH emission appears approximately co-spatial with the H$_2$CO emission. The perfect kinematic analogy between the two lines also suggest that the two types of emission are, to first order, spatially coincident.

\subsection{Column densities in the midplane and molecular layer}
\label{column_densities}

The column densities of the detected molecular species were inferred by assuming local thermodynamic equilibrium and optically thin lines. The validity of these assumptions is discussed by \citet{podio19}. The CS, H$_2$CO, and CH$_3$OH line intensities were integrated over the following two distinct regions: the disk molecular layer, as defined in Sect. \ref{sect:methanol} ($A=4.3$ arcsec$^2$),
%defined as the X-shaped region where CS and H$_2$CO emission is brighter ($A=4.3 \arcsec^2$), 
and the outer disk mid-plane ($A=0.41$ arcsec$^2$) (see sketch in Fig.\,\ref{fig:overview}). The noise on the integrated emissions, $F_{\rm int}$, was derived just as for methanol\textsuperscript{\ref{note1}}. If $F_{\rm int} < 3\sigma$, we report the upper limit (Table \ref{tab:column_densities}).
Then, the integrated line intensities were converted into column densities through the molecular parameters in Table \ref{tab:lines}  \citep[CDMS,][]{Mueller2005}. 
As there are no estimates of the disk temperature, the excitation temperature is assumed to be 20$-$100 K in the molecular layer and 10$-$30 K in the outer midplane \citep[see e.g.,][]{walsh10}. The minimum and maximum column densities over this range of $T_{\rm ex}$ are summarized in Table \ref{tab:column_densities}. The assumed $T_{\rm ex}$ values are in agreement with those inferred from multi-line observations of CS and H$_2$CO in a few Class II disks \citep[e.g., ][]{legal19,pegues20}, while no estimates of $T_{\rm ex}$ in Class I disks are available. The total H$_2$CO and CH$_3$OH column densities were derived assuming an ortho-to-para ratio of $1.8-2.8$ \citep{guzman18a} and the ratio of A-type to E-type forms to be one. 

The average H$_2$CO column density in the outer disk midplane is up to a factor of 8 lower than in the molecular layer, whereas the CS column density decreases by a factor of up to $>10$ in the midplane, which is in agreement with the decrease in line intensity shown by the vertical profiles in Fig.~\ref{fig:vertical-profiles}. As for CH$_3$OH, the column density is determined in the molecular layer, while only an upper limit is given for the outer midplane. 
%From the o-H$_2$CO column density, we inferred the total H$_2$CO column density by assuming an ortho-to-para ratio of $1.8-2.8$ \citep{guzman18a}. %This turned out to be $N_{\rm H_2CO} \sim 7-25 \times 10^{13}$ cm$^{-2}$ in the molecular layer and $\sim 2.8-5.6 \times 10^{13}$ cm$^{-2}$ in the midplane.
From the inferred values, the abundance ratio between CH$_3$OH and H$_2$CO is $\sim 0.5-0.6$ in the molecular layer and $< 3.2$ in the midplane.

\begin{table}
  \caption[]{\label{tab:column_densities} Integrated intensities and average column densities of CS, H$_2$CO, and CH$_3$OH calculated over the molecular layer ($T_{\rm ex} = 20-100$ K) and outer midplane ($T_{\rm ex} = 10-30$ K) (see Sect.\,\ref{column_densities}).
  %The brackets denote integrated intensity at 2$\sigma$.
  }
  \centering
  \begin{tabular}[h]{cccc}
    \hline
    \hline
  Species & Disk region & $F_{\rm int}$ & $N$ \\
   &  & (mJy km/s) & (10$^{13}$ cm$^{-2}$) \\
   \hline
\multirow{2}{*}{CS} & Molecular layer & 2134 (31) & 3.4$-$5.0 \\
  \smallskip
 & Outer midplane & $< 30$ & $< 1.8$, $< 0.5$ \\
  \multirow{2}{*}{H$_2$CO} & Molecular layer & 1865 (26) & 7.2$-$25 \\
  \smallskip
   & Outer midplane & 72 (8) & 2.8$-$5.6 \\
  \multirow{2}{*}{CH$_3$OH} & Molecular layer & 42 (13) & 3.6$-$14.6 \\
  \smallskip
  & Outer midplane & $< 12$ & $< 18$ \\
 \hline     
  \end{tabular}
  \tablefoot{The integrated intensities (errors in brackets) refer to the CS $5-4$, o-H$_2$CO $3_{1,2}-2_{1,1}$, and CH$_3$OH $5_{0,5}-4_{0,4}$ (A) lines.}
\end{table}

%\begin{figure*}
%  \centering
% \includegraphics[width=13cm]{DALI-lines.png} 
%     \caption{Molecular abundances and line emitting region as predicted by DALI.} 
% \label{fig:dali}
% \end{figure*}

\section{Discussion}

%    \item[-]
The observed molecular emission highlights the disk's vertical stratification.
The CO $2-1$ emission probes an upper disk layer ($z/r \sim 0.41-0.45$), which extends up to a larger disk height with respect to o-H$_2$CO $3_{1,2}-2_{1,1}$ and CS $5-4$ ($z/r \sim 0.2-0.3$).
The H$_2$CO and CS emissions are roughly coincident and trace an intermediate disk layer. The analogous distribution of the two lines is explained by (i) the very similar upper level energy and critical densities ($E_{\rm up}=33$ K, and 35 K, and $n_{\rm cr} \sim 7-5 \times 10^5$ cm$^{-3}$, and $17-9.5 \times 10^5$ cm$^{-3}$, at $20-100$ K, for the H$_2$CO and the CS line, respectively, \citealt{shirley15}) and (ii) the similar distribution of the H$_2$CO and CS abundance in the disk suggested by thermo-chemical models \citep[e.g., ][]{fedele20}. Both molecules are released from grains and/or formed in the gas phase in the molecular layer, following the photodissociation of CO, which makes available C for the formation of small hydrocarbons, together with atomic O. The CS is formed from reactions of small hydrocarbons with either ionized or atomic S \citep[e.g., ][]{legal19}, while the main formation route of H$_2$CO is via the reaction CH$_3$+O \citep[e.g., ][]{loomis15}. 
    %and they have common formation and destruction paths.    The chemical models by \citet{legal19} suggests that the main formation routes of CS are: either via rapid ion–neutral reactions between S$^+$ and small hydrocarbons (such as CH$_x$ and C$_y$H), which produces carbonated S-ions, including HCS$^+$, CS$^+$, HC3S$^+$, and C2S$^+$, that subsequently recombine with electrons to form neutral S-bearing species; or via  neutral–neutral reactions between S and small hydrocarbons (at deeper disk layers). The main formation route of H$_2$CO in gas phase is through the reaction CH$_3$ + O, which is efficient in the warm inner region and upper layers of the disk where atomic oxygen is produced by photodissociation of gas-phase CO. The latter, also makes available C for the formation of small hydrocarbons which boost the formation of both H$_2$CO and CS. Finally, both species are easily destructed in the disk atmosphere due to photo-dissociation and  reactions with protons and protonated ions (i.e., with H$^+$, H$_3^+$, and HCO$^+$, and will freeze-out onto dust grains in the disk midplane. Therefore, if H$_2$CO and CS are mainly formed in gas-phase in the disk molecular layer, the abundance of the two molecules would be strongly linked to the presence of small hydrocarbons.\\
%  \item[-]
Additionally, CO $2-1$ and CS $5-4$ emissions strongly decrease in the outer disk-midplane, where molecules freeze-out onto dust grains (freeze-out layer). By contrast, H$_2$CO is only a factor of 2 less intense, which indicates that the H$_2$CO molecules trapped on the icy grains in the cold midplane are partially released by nonthermal processes (UV, X-ray, cosmic-rays-induced, and/or reactive desorption).
We tentatively detected CH$_3$OH emission at $3\sigma$ in the disk region where H$_2$CO and CS emissions are bright, that is to say the warm molecular layer where molecules should be efficiently released from grains, while it remains undetected in the outer disk midplane.
In the right panel of Fig. \ref{fig:ch3oh}, the [CH$_3$OH]/[H$_2$CO] abundance ratio inferred for the Class I disk of IRAS 04302 is compared with the estimates obtained in hot corinos around Class 0 protostars, in the other Class I disks observed by ALMA-DOT, in the Class II disks of TW Hya and HD 163296, and in comets. The abundance ratio in the disk of IRAS 04302 ($\sim 0.5-0.6$ for T$_{\rm ex} \sim 20-100$ K) is consistent with the upper limits derived for the other Class I and early Class II disks observed by ALMA-DOT (between $<0.7$ and $<3.2$, \citealt{podio19,garufi20b}, Garufi et al. in prep.).
On the other hand, it is lower by a factor of $\sim2.5$ than that inferred for the Class II disk of TW Hya ($\sim 1.27 - 1.73$ for T$_{\rm ex} \sim 25-75$ K, \citealt{walsh16,carney19}), while it is larger than the upper limit obtained for HD 163296 ($<0.24$, \citealt{carney19}). As methanol forms on the dust grains due to CO freeze-out and subsequent hydrogenation, the lower [CH$_3$OH]/[H$_2$CO] in HD 163296 may be due to a smaller degree of CO ice chemistry in the warmer disks around Herbig stars \citep[e.g., ][]{pegues20}. %, as suggested by \citet{pegues20} based on lower H$_2$CO column densities in the Herbig disks in their sample. 
The abundance ratios in Class I and Class II disks are from a factor of $2$ to 2 orders of magnitude lower than the values inferred in the hot corinos around the Class 0 protostars IRAS 16293-2422 A and B in Taurus \citep{jorgensen18,persson18,manigand20}, and NGC1333-IRAS 4A and IRAS 2A in Perseus \citep{taquet15}. This difference may be due to an evolution of the methanol and formaldehyde abundance, which would indicate that the chemical processes occurring in the disk alter its chemical composition with respect to the protostellar stage ("disk reset" scenario). If instead the chemical composition of the ices in the disk is inherited from the protostellar stage and remains unaltered ("inheritance" scenario), the observed difference may either depend on the fact that the emission around Class 0 sources is unresolved or on the different processes which release the molecules in the gas phase. While, in the hot-corino, the ices are released in the gas phase by thermal desorption at $T > 100$ K \citep[e.g., ][]{ceccarelli07}, in the disk midplane and molecular layer, nonthermal processes are at play (photo-desorption, cosmic-rays-induced, and/or reactive desorption, e.g., \citealt{walsh14}). A detailed comparison with thermo-chemical disk models is necessary to distinguish between the two competing scenarios. Finally, the [CH$_3$OH]/[H$_2$CO] abundance ratio in all disks except HD 163296 is comparable with the wide range found for comets 67P/C-G, Hale-Bopp, and Lovejoy ($\sim 0.65-8$, \citealt{biver15,rubin19}), even though IRAS 04302 sits at the lower limit of that range. The comparison in Fig. \ref{fig:ch3oh} suggests a change in the chemical composition in disks with respect to the protostellar stage, which is then inherited by forming planets and minor bodies, such as comets. 
%is compared with what is found in comets, which are believed to keep track of the chemical composition of the protosolar nebula. The abundance ratio in all disks except HD 163296 are comparable with the wide range found for comets 67P/C-G, Hale-Bopp, and Lovejoy, even though IRAS 04302 sits at the lower limit of that range ([CH$_3$OH]/[H$_2$CO]$\sim 0.65-8$, \citealt{biver15,rubin19}). 
%Recent studies suggest that the abundance of COMs in Class 0 and I hot-corinos is comparable with that of comets \citep{bianchi19a,drozdovskaya19}, supporting the inheritance scenario.  
 %This is consistent with the idea that ice chemistry should be more efficient in the cold midplane of disks around T Tauri stars with respect to strongly irradiated disks around Herbig stars.    

\section{Conclusions}

The nearly edge-on disk around the Class I source IRAS 04302 observed at $\sim 48$ au resolution in the context of the ALMA-DOT campaign allows us to resolve, for the first time, the vertical structure of a protoplanetary disk in several molecular tracers. The CO, CS, and H$_2$CO maps probe the chemical stratification predicted by thermo-chemical models \citep[e.g.,][and references therein]{dutrey14}: the warm molecular layer, with the height of the emitting region linearly increasing with the radial distance; and the freeze-out layer in the cold outer midplane where line emission decreases due to molecular freeze-out onto dust grains.

We report the first tentative detection of CH$_3$OH in the disk of a Class I source and constrain the abundance with respect to H$_2$CO to be 0.5$-$0.6. The inferred abundance ratio is 1-2 orders of magnitude lower than in hot corinos around Class 0 protostars, while it is only a factor of $2.5$ lower than that estimated in the Class II disk of TW Hya and comparable with that inferred for comet 67P/C-G. This may be due to chemical evolution, that is, to a chemical reset in the disk with respect to the protostellar stage and/or to the different processes responsible for the release of molecules in the gas phase (thermal desorption in hot corinos and nonthermal desorption in disks). A detailed modeling of the disk thermo-chemical structure is required to distinguish between inheritance and disk-reset scenarios. %Moreover, resolved observations of molecular emission in the disk at all stages from Class 0 to Class II are needed to properly compute the abundance ratios, and trace their chemical evolution.    

\begin{acknowledgements}
We thank C.\ Spingola for the valuable help with the ALMA data reduction. This paper makes use of the ALMA data 2018.1.01037.S (PI L. Podio). ALMA is a partnership of ESO (representing
its member states), NSF (USA) and NINS (Japan), together with NRC
(Canada), MOST and ASIAA (Taiwan), and KASI (Republic of Korea), in cooperation with the Republic of Chile. 
This work was supported  
by:  (i) the EU's Horizon 2020 research and innovation programme under Marie Sk{\l}odowska-Curie grant agreement No  811312 (Astro-Chemical Origins), and No 823823 (RISE DUSTBUSTERS); (ii) the ERC project "The Dawn of Organic Chemistry" (DOC), grant agreement No
741002; (iii) PRIN-INAF/2016 GENESIS-SKA; (iv) the Italian Ministero dell’Istruzione, Universit{\'a} e Ricerca through the grants Progetti Premiali 2012/iALMA (CUP-C52I13000140001), 2017/FRONTIERA (CUP-C61I15000000001), and SIR-(RBSI14ZRHR); (v) the Deutsche Forschungs-gemeinschaft (DFG) - Ref no. FOR2634/1TE1024/1-1; (vi) the DFG cluster of excellence Origins (www.origins-cluster.de); (vii) the French National Research Agency in the framework of the Investissements d’Avenir program (ANR-15-IDEX-02), through the "Origin of Life" project of the Univ. Grenoble-Alpes.
\end{acknowledgements}

\bibliographystyle{aa} % style aa.bst 
\bibliography{mybibtex.bib} % your references Yourfile.bib

%\begin{thebibliography}{}
%\bibitem[1997]{zheng} Zheng, W., Davidsen, A. F., Tytler, D. \& Kriss, G. A.      1997, preprint
%\end{thebibliography}
%
%%%%%%%%%%%%%%%%%%%%%%%%%%%%%%%%%%%%%%%%%%%%%%%%%%%%%%%%%%%%%%

%
%-------------------------------------------------------------
%               Appendices have to be placed at the end, after
%                                        \end{thebibliography}
%-------------------------------------------------------------

\begin{appendix} %First appendix
\section{Properties of the observed lines}

In Table \ref{tab:lines}, the properties of the observed lines (species, transition, frequency in MHz, upper level energy (E$_{\rm up}$) in K, and line strength ($S_{ij} \mu^2$) in D$^2$) and of the obtained line cubes and moment 0 maps (clean beam in $\arcsec \times \arcsec$, channel width in km s$^{-1}$, r.m.s noise per channel in mJy beam$^{-1}$,  and r.m.s of the moment 0 map, $\sigma_0$, in mJy beam$^{-1}$ km s$^{-1}$) are summarized.

\begin{table*}
\centering
  \caption[]{\label{tab:lines} Properties of the observed lines from the CDMS database \citep{Mueller2005} and of the obtained line cubes and moment 0 maps.}
  \tiny{
  \begin{tabular}[h]{ccccccccc}
    \hline
    \hline
 Species & Transition & Frequency & $E_{\rm up}$ & $S_{ij} \mu^2$ & Clean beam & Channel width & r.m.s. & $\sigma_0$    \\
         &            & (MHz)     & (K)          &    (D$^2$)     & ($\arcsec\times\arcsec$) & (km s$^{-1}$) & (mJy beam$^{-1}$) & (mJy beam$^{-1}$ km s$^{-1}$) \\
    \hline 
\smallskip
CO & $2-1$ & 230538.000 & 17 & 0.02 & 0.33$\times$0.25 & 0.2 & 2.3 & 9 \\ 
\multirow{3}{*}{CN$^a$} & $2-1$, J=5/2-3/2, F=5/2-3/2   & 226874.1908 & 16 & 4.2   &  &  &  \\
& $2-1$, J=5/2-3/2, F=7/2-3/2 & 226874.7813 & 16 & 6.7   & 0.34$\times$0.26    &  0.2    & 2.0 & 4.1   \\
\smallskip
& $2-1$, J=5/2-3/2, F=3/2-1/2 & 226875.8960 &16 & 2.5   &  &      &       \\
\smallskip
CS & $5-4$                             & 244935.557  & 35 & 19.1 & 0.31$\times$0.26 & 1.2  & 0.8 & 4.6 \\
\smallskip
o-H$_2$CO & $3_{1,2}-2_{1,1}$          & 225697.775  & 33 & 43.5 & 0.34$\times$0.26  & 0.2 & 2.0 & 4.2\\
CH$_3$OH-A & $5_{0,5}-4_{0,4}$         & 241791.352  & 35 & 16.2 & 0.41$\times$0.32  & 0.2 & 1.7 & 2.3\\
%\smallskip
%\multirow{2}{*}{CH$_3$OH-A$^b$} & \multirow{2}{*}{$5_{0,5}-4_{0,4}$} & \multirow{2}{*}{241791.352} & \multirow{2}{*}{35} & \multirow{2}{*}{16.2} & 0.32$\times$0.24 & 0.2 & 2.1 \\
%\smallskip
% & & & & & 0.41$\times$0.32 & 0.2 & 1.7 \\
    \hline     
  \end{tabular}
}
%\small
%\smallskip
\tablefoot{($^{a}$) The CN $2-1$ transition consists of 19 hyperfine structure components. The ALMA SPW is centered on the second component listed in the table, which is the brightest hyperfine component of CN $2-1$. Because of the line broadening due to disk kinematics, this is blended with the other two listed components. 
%The integrated line intensity refer to the sum of the blended components.
%($^{b}$)
%Methanol occurs as A− and E− symmetry states. The ratio of A-type to E-type forms of methanol is one.
%The two beam sizes listed for the CH$_3$OH line correspond to the two alternative reductions performed  using Briggs  weighting  with  robust=0.0, and 2.0 (see Sect.\,\ref{data_reduction})}.
}
\end{table*}

\section{Moment 1 maps of line emission}

\begin{figure*}
  \centering
 \hspace{-1.5cm}
 \includegraphics[width=8.2cm]{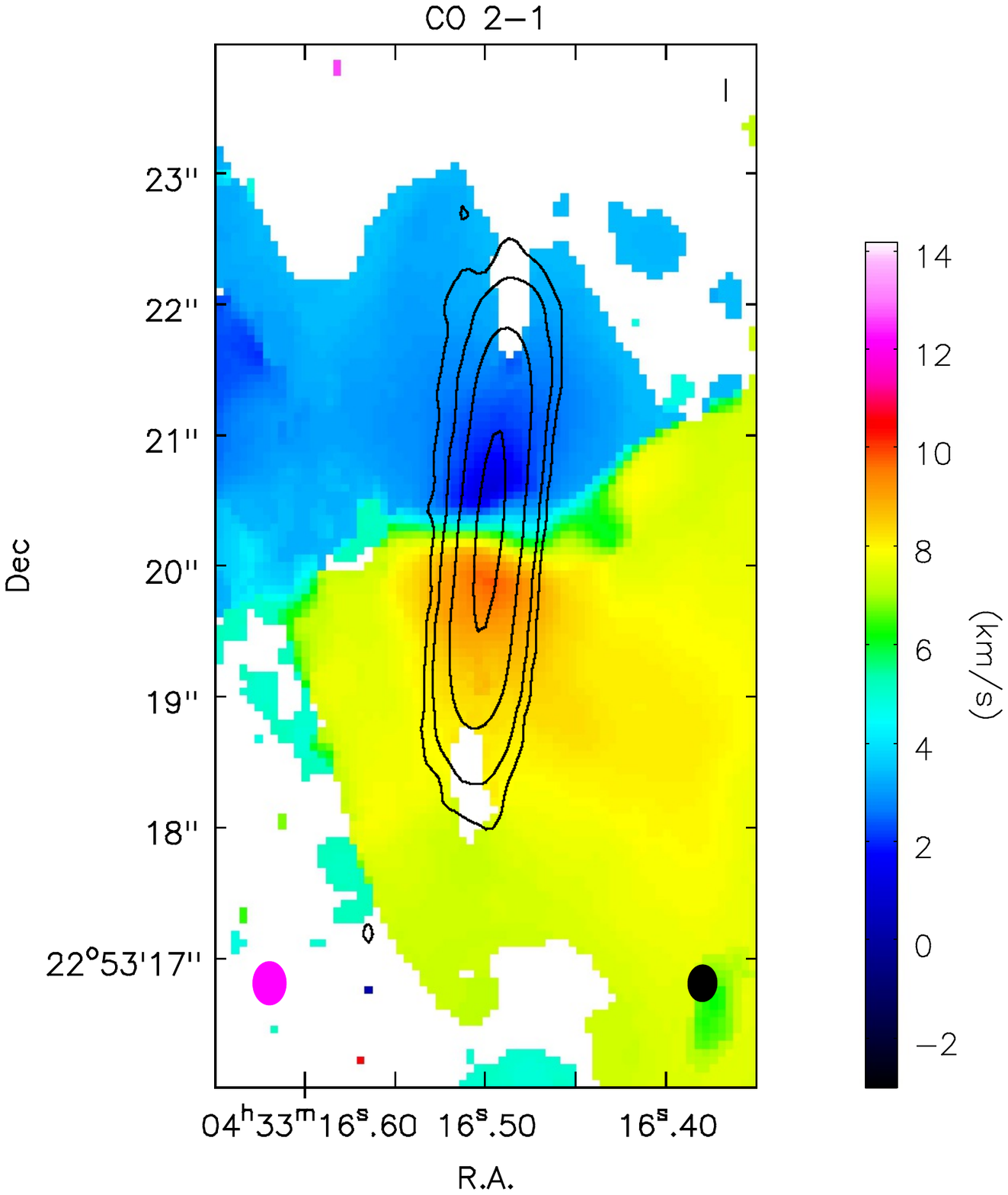} 
 \hspace{-1.8cm}
 \includegraphics[width=8.2cm]{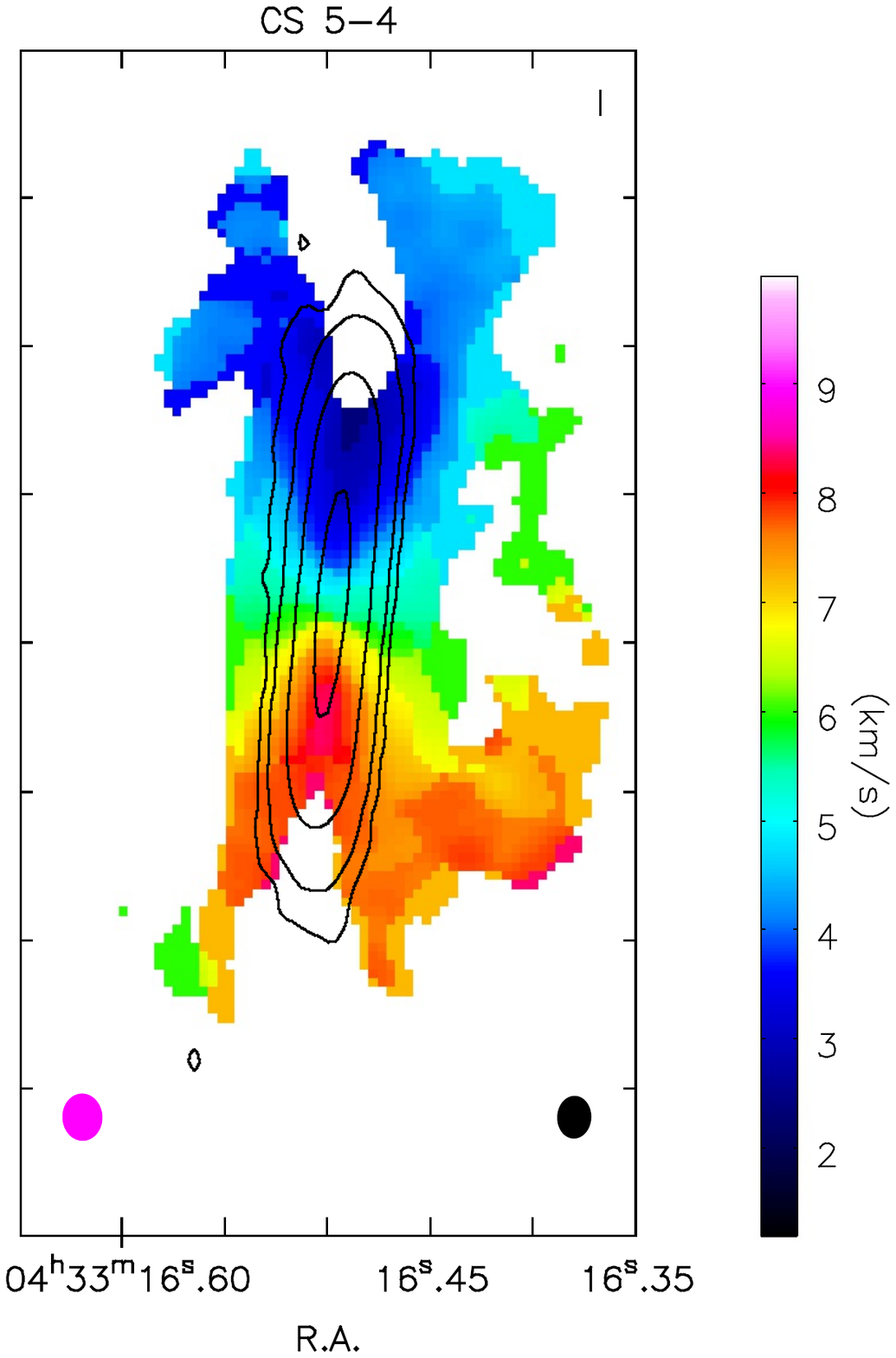} 
 \hspace{-3.4cm}
 \includegraphics[width=8.2cm]{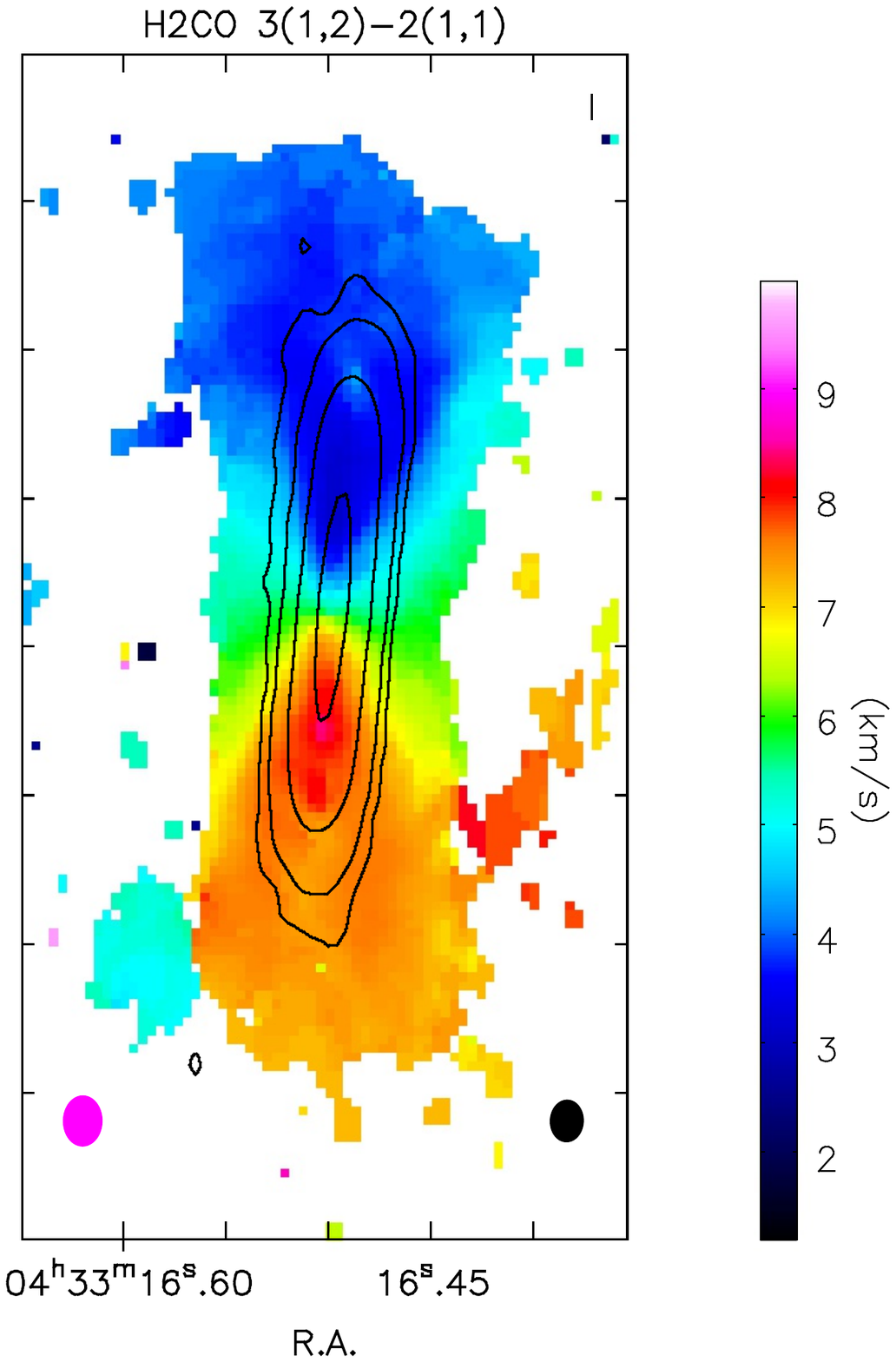}
%\hspace{-4.cm}
%\includegraphics[width=5.5cm]{H2CS_mom1.pdf} 
 %\vspace{-3cm}
     \caption{Moment 1 maps of CO $2-1$ (velocity scale from $-3$ to $+14.2$ km s$^{-1}$), CS $5-4$, and H$_2$CO $3_{1,2}-2_{1,1}$ (velocity scale from $+1.2$ to $+10$ km s$^{-1}$). Black contours indicate the continuum emission at 1.3~mm ($3\sigma$, $10\sigma$, $50\sigma$, and $200\sigma$ intensity levels, with $\sigma = 0.05$ mJy beam$^{-1}$). The magenta (black) ellipses on the bottom-left (right) corners show the line (continuum) beam.} 
 \label{fig:mom1}
 \end{figure*}

Figure \ref{fig:mom1} shows the intensity weighted velocity distributions (moment 1 maps) of CO $2-1$, CS $5-4$, and o-H$_2$CO $3_{1,2}-2_{1,1}$ obtained by applying a $3 \sigma$ clipping. The contours of the continuum emission at 1.3~mm are overplotted.  The moment 1 maps show the typical disk rotation pattern. Even if the H$_2$CO line intensity decreases in the inner 48 au and in the outer disk midplane (see the moment 0 map in Fig. \ref{fig:overview} and vertical profiles in Fig. \ref{fig:vertical-profiles}), the moment 1 map shows that the H$_2$CO emission probes the gas rotation in the disk from the inner region out to a radius of $\sim 3\arcsec$ ($R_{\rm H_2CO} \sim 480$ au). The H$_2$CO emission is, therefore, more extended than the continuum by a factor of $\sim 1.4$ (R$_{\rm dust} \sim 350$ au as determined from the $3\sigma$ contour level). A similar velocity distribution is shown by CS $5-4$ and CO $2-1$, except that no emission at $> 3\sigma$ is detected in the outer disk midplane, that is to say for radii $\ge 1.4\arcsec$ corresponding to $\sim230$ au, as is also shown by the vertical profile at this radius (Fig. \ref{fig:vertical-profiles}). Finally the moment 1 map of CO $2-1$ also shows the velocity distribution of the gas in the envelope, as the emission in this line extends well beyond the gaseous disk structure probed by the o-H$_2$CO $3_{1,2}-2_{1,1}$ line.

%\end{comment}

\end{appendix}
\end{document}